\documentclass[reprint,nofootinbib,amsmath,amssymb,fontenc aps]{revtex4-1}
\usepackage{graphicx,dcolumn,bm,hyperref,float}
\usepackage{anyfontsize}
\usepackage{color}
\usepackage{xcolor}

\newcommand{\beq}{\begin{equation}}
\newcommand{\eeq}{\end{equation}}
\newcommand{\bea}{\begin{eqnarray}}
\newcommand{\eea}{\end{eqnarray}}
\newcommand{\bef}{\begin{figure}}
\newcommand{\eef}{\end{figure}}

\newcommand{\mpl}{M_{\mbox{\tiny{Pl}}}}

\newcommand{\co}{b}
\newcommand{\cor}{a}
\newcommand{\corr}{c}
\newcommand{\pol}{\varepsilon}

\begin{document}

\title{\fontsize{11.9}{15}\selectfont Solar System Constraints on Light Propagation from Higher Derivative \\ Corrections to General Relativity and Implications for Quantum Gravity}
\title{\fontsize{11.9}{15}\selectfont Solar System Constraints on Light Propagation from Higher Derivative \\ Corrections to General Relativity and Implications for Fundamental Physics}

\author{Mark P.~Hertzberg$^{1,2,3}$}
\email{mark.hertzberg@tufts.edu}
\author{Rachel Nathan$^1$}
\email{rachel.nathan@tufts.edu}
\author{Suzanna E.~Semaan$^{4,1}$}
\email{ssemaan@brynmawr.edu}
\affiliation{$^1$Institute of Cosmology, Department of Physics and Astronomy, Tufts University, Medford, MA 02155, USA
\looseness=-1}
\affiliation{$^2$Institute of Theory and Computation, Center for Astrophysics, Harvard University, 60 Garden Street, Cambridge, MA 02138, USA
\looseness=-1}
\affiliation{$^3$Center for Theoretical Physics, Department of Physics, 
Massachusetts Institute of Technology, Cambridge, MA 02139, USA}
\affiliation{$^4$Department of Physics and Mathematics, Bryn Mawr College, Bryn Mawr, PA 19010, USA
\looseness=-1}

\begin{abstract}
While the two derivative action of gravitation is specified uniquely, higher derivative operators are also allowed with coefficients that are not specified uniquely by effective field theory. We focus on a four derivative operator in which the Riemann tensor couples directly to the electromagnetic field  $\cor\,R_{\mu\nu\alpha\beta}F^{\mu\nu}F^{\alpha\beta}$. We compute the corresponding corrections to the Shapiro time delay in the solar system and compare this to data from the Cassini probe. We place an observational upper bound on the coefficient $\cor$ at 95\% confidence $|\cor|<26\,(1000\,\mbox{km})^2$. By way of motivation, we also compare this to a weak gravity conjecture (WGC) prediction of a bound on the coefficients $\cor,\,\co$ of four derivative operators involving the graviton and the photon; this includes the above term $\cor\,R_{\mu\nu\alpha\beta}F^{\mu\nu}F^{\alpha\beta}$ as well as $\co\,F^4$. We show that by using the observed value of the $\co$ coefficient from measurements of light by light scattering, which arises in the Standard Model from integrating out the electron, the WGC predicted bound for $\cor$ is $\cor\lesssim 7.8\,(1000\,\mbox{km})^2$. This is consistent with the above observational bound, but is intriguingly close and can be further probed in other observations.
\end{abstract}

\maketitle


\section{Introduction}

General relativity is known to be a unique theory in a precise sense: it is the unique theory of massless spin 2 particles at large distances. More precisely, this means the action is uniquely the Einstein-Hilbert action at the {\em two derivative} level. However, if one consider higher derivatives, then corrections are allowed. In fact corrections are inevitable since the theory is non-renormalizable, so such corrections are generated even if not included in the starting classical action. Furthermore, such corrections are predicted by UV completions, such as string theory.

From the point of view of effective field theory (EFT), the higher order operators have coefficients that are left unspecified. However, from a range of considerations in recent years, including consistency with causality and the weak gravity conjecture, these coefficients are understood to not be completely arbitrary but obey certain inequalities. A known inequality in the literature relates the coefficients of operators quartic in the electromagnetic fields to the coefficient of operators that couple the electromagnetic field to gravity. 

Normally, these inequalities are thought to be robust but very difficult to test experimentally. The most well known inequality from the weak gravity conjecture (WGC) \cite{Arkani-Hamed:2006emk} is $m\leq \sqrt{2}\,e\,\mpl$, where $m$ is the mass of the lightest charged state, $e$ is the electric coupling, and $\mpl=1/\sqrt{8\pi G_N}$ is the Planck mass. If we apply this to the electron with $m_e\approx 0.5\,$MeV, electric coupling $e\approx 0.3$ and Planck mass $\mpl\approx 2.4\times 10^{18}\,$GeV, this inequality is satisfied by many orders of magnitude. By contrast, the weak gravity conjecture inequality that we will focus on in this work is surprisingly {\em testable}, as we will explain. This will come with the very large caveat that to be near the inequality, the cut off on the effective theory will be very low, requiring new physics to enter at very low scales. While such low scale UV cut offs do not sound (to us) very plausible, we think it is still worth subjecting any claims from quantum gravity to the test, regardless of how weak the tests may be. 

The WGC bound that we will discuss will impose an inequality on the coefficient $\cor$ of the four derivative operator $a\,R_{\mu\nu\alpha\beta}F^{\mu\nu}F^{\alpha\beta}$ in terms of the coefficient $\co$ of the four derivative operator $\co\,F^4$. Since the coefficient $\co$ has been measured in light by light scattering observations, we can then obtain a numerical bound on $\co$, which turns out to be $a\lesssim 7.8\,(1000\,\mbox{km})^2$ (the reason for the imprecision will be discussed later). Interesting if nature is near this bound for $\cor$, this gives rise to a correction to the propagation of light in the solar system in the range of detectability. We compute the correction to the Shapiro time delay in the solar system and apply it to the Cassini probe. We perform a detailed analysis of the data from  Cassini, finding an observational bound $|\cor|<26\,(1000\,\mbox{km})^2$, which is rather close to the WGC conjecture bound. 

The outline of our paper is as follows: 
In Section \ref{Action} we present the class of theories we are considering.
In Section \ref{Measurements} we discuss some existing measurements of coefficients and a WGC bound on others.
In Section \ref{Altered} we compute the corrections to the Shapiro time delay.
In Section \ref{Data} we recap the Cassini probe data and compare to this theory.
In Section \ref{Likelihood} we perform a statistical analysis and report on our findings.
Finally, in Section \ref{Discussion} we provide further discussion.

\section{Action}\label{Action}

The standard action that couples electromagnetism to gravity is (units $\hbar=c=1$)
\beq
S_2=\int d^4x\sqrt{-g}\left[{1\over 2}\mpl^2\,R-{1\over 4}F_{\mu\nu}F^{\mu\nu}\right],
\eeq
where $R$ is the Ricci scalar and $F_{\mu\nu}$ is the electromagnetic field strength.  This action is the {\em unique} action at the {\em two derivative} level (hence we labelled it $S_2$).  However, there is no a priori reason there should not be higher dimension operators. In fact there are multiple reasons they {\em must} be there: (i) since the above action is non-renormalizable, they are generated under renormalization, (ii) if one integrates out massive charged particles, like the electron, they are generated. 

Let us consider the most general four derivative correction to the action (we label it $S_4$) involving both the graviton and the photon. We write this as
\bea
S_4 = \int d^4x\sqrt{-g}\Big{[}
\co\,(F^2)^2+\co'\,F^4+\co'' D_\mu F^{\mu\alpha}D_\nu F^{\nu}_{\,\,\alpha}\nonumber\\
-\cor\, R_{\mu\nu\alpha\beta}F^{\mu\nu}F^{\alpha\beta}-\cor'R_{\mu\nu}F^{\mu\alpha}F^\nu_{\,\,\alpha}-\cor''RF^2\nonumber\\
+\corr\,R_{\mu\nu\alpha\beta}R^{\mu\nu\alpha\beta}+\corr'\,R_{\mu\nu}R^{\mu\nu}+\corr''\,R^2
\Big{]},
\eea
where $F^2\equiv F_{\mu\nu}F^{\mu\nu}$ and $F^4\equiv F_{\mu\nu}F_{\alpha\beta}F^{\mu\alpha}F^{\nu\beta}$. 
We have labelled the coefficients of these terms as $\cor,\,\cor',\,\cor'',\,\co,\,\co',\,\co'',\,\corr,\,\corr',\,\corr''$. 
One can also include two other four derivative terms, namely $D_\nu F^{\mu\alpha} D_\mu F^{\nu}_{\,\,\alpha}$ and $D_\alpha F^{\mu\nu}D^\alpha F_{\mu\nu}$; however, by integrating by parts and using some identities, their effects can be absorbed into the terms included here without loss of generality.
In this work, we are interested in constraining the above coefficients. For reasons we will explain, the coefficient $\cor$ will play a central role in our work.

\section{Measurements and \\ Weak Gravity Bound}\label{Measurements}

The coefficients of these higher dimension operators are just parameters of the effective field theory. However, we can constrain them. Within the Standard Model, one can integrate out charged particles and several are generated at 1-loop, with the largest contribution coming from the electron. 

If the value at distances shorter than the Compton wavelength of the electron is denoted $\cor_{\mbox{\tiny{UV}}},\,\co_{\mbox{\tiny{UV}}}$, then for larger length scales, the values of the $\co$ coefficients are shifted at 1-loop to \cite{Heisenberg:1936nmg}
\bea
&&\co=-{\alpha^2\over 36\,m_e^4}+\co_{\mbox{\tiny{UV}}},\\
&&\co'=+{7\alpha^2\over 90\,m_e^4}+\co'_{\mbox{\tiny{UV}}},
\eea
where $m_e\approx0.5$\,MeV is the electron mass and $\alpha\approx1/137$ is the fine structure constant.
The $\sim (F^2)^2$ and $\sim F^4$ interactions have been detected by measurements of light by light scattering at the LHC \cite{ATLAS:2017fur,TOTEM:2021zxa}, with consistency with the values generated by charged particles in the Standard Model. So for them we can safely ignore any $\co_{\mbox{\tiny{UV}}}$ contributions at low energies.
So we have 
\beq
\co\approx -{\alpha^2\over 36\,m_e^4},\,\,\,\,\,\, \co'\approx {7\alpha^2\over 90\,m_e^4}.
\eeq

Integrating out the electron also generates the $\sim R_{\mu\nu\alpha\beta}F^{\mu\nu}F^{\alpha\beta},\,R_{\mu\nu}F^{\mu\alpha}F^\nu_{\,\,\alpha},\,R F^2$ terms with shifted values $\Delta \cor,\Delta\cor',\,\Delta\cor''\sim \alpha/m_e^2$, as computed in Ref.~\cite{Drummond:1979pp} (updates include Ref.~\cite{Bastianelli:2008cu}). If this is the correct value for $\cor,\,\cor',\,\cor''$, then it is easy to check that their effects are so small in the solar system and are essentially unmeasurable. However, since we currently do not have a measurement of these coefficients, we do not know that this is the correct value. All we know is that the value of $\cor,\,\cor',\,\cor''$ will {\em change}, or ``flow", by $\Delta \cor,\,\Delta\cor',\,\Delta\cor''$ above and below the electron mass. It could be that their actual values are much larger than this (positive or negative), and the correction from integrating out the electron is a negligible correction. So for the purpose of this paper, we shall consider the possibility that $\cor,\,\cor',\,\cor''$ are large as provided by their UV values. So we shall take 
\beq
\cor\approx\cor_{\mbox{\tiny{UV}}},\,\,\,\cor'\approx\cor'_{\mbox{\tiny{UV}}},\,\,\,\cor''\approx\cor''_{\mbox{\tiny{UV}}}
\eeq
as free parameters, which we wish to constrain both theoretically and observationally.

There are theoretical reasons to believe that the values of these coefficients in the effective field theory are not completely arbitrary.
A very interesting bound comes from the WGC. Here one considers the above four derivative action and computes the corrections to the charge-mass ratio of an extremal black hole of mass $M$ and charge $Q$. In general relativity, an extremal black hole has $M=\sqrt{2}\,|Q|\mpl$. With these higher order terms, the leading correction is found to be \cite{Kats:2006xp}
\beq
{M\over \sqrt{2}\,|Q|\mpl}=1-\delta,
\eeq
with 
\beq
\delta={64\pi^2\over 5 Q^2}\left(4\corr+\corr'-\mpl^2(\cor+\cor')+\mpl^4(4\co+2\co')\right).
\eeq
In order to maintain the WGC, one needs this leading correction to obey the inequality 
\beq
\delta\geq0,
\eeq
so that $M\leq\sqrt{2}\,|Q|\mpl$ for extremal black holes, just as elementary particles should obey $m\leq\sqrt{2}\,e\,\mpl$ in the original form of the WGC.
We note that other interesting developments include Refs.~\cite{Cheung:2014ega,Cremonini:2009ih,Bellazzini:2019xts,Cremonini:2019wdk,Noumi:2022ybv,BarrosoVarela:2023ull,Abe:2023anf,Bittar:2024xuc,Barbosa:2025uau}.

We note that since this WGC makes use of charged black holes, which have $R=0$ in vacuum (due to conformal symmetry), all operators involving the Ricci scalar $R$ in the action $S_4$ play no role; so this is why $\cor''$ and $\corr''$ do not appear in $\delta$. Also the term involving $D_\mu F^{\mu\alpha}$ is negligible, as it vanishes to leading order in vacuum by the Maxwell equations; so this is why $\co''$ does not appear in $\delta$ either. 

The coefficients $\corr,\,\corr'$ of the pure gravity terms $R_{\mu\nu\alpha}R^{\mu\nu\alpha\beta},\,R_{\mu\nu}R^{\mu\nu}$ are dimensionless  and have no a priori reason to be large. We will assume they are not extremely large; which renders them negligible in the above inequality, as they do not multiply powers of $\mpl$ in $\delta$ as the other terms do. Hence the $\delta\geq0$ WGC becomes
\beq
\cor+\cor'\leq (4\co+2\co')\mpl^2.
\eeq
By inserting the above values of $\co$ and $\co'$ into this, we have the theoretical bound
\beq
\cor+\cor' \leq {2\,\alpha^2\mpl^2\over 45\,m_e^4}.
\label{WGCbound}\eeq
Evaluating this with the known values of $\alpha,\,m_e,\,\mpl$ we obtain a form of the WGC bound as
\beq
\cor +\cor' \leq 7.8\, (1000\,\mbox{km})^2,
\eeq
where we have expressed the answer in units of $(1000\,\mbox{km})^2$ for convenience. 
If $\cor,\cor'$ are very small, such as the value $\Delta\cor\sim \alpha/m_e^2$ from the shift provided by the electron, they easily satisfy this bound. 

If we allow $\cor+\cor'$ to be positive and large, then this bound is quite interesting and perhaps surprising. It says that considerations of the WGC, which is often thought to be very difficult to test, yields a value for the characteristic length scale that controls the $\sim R_{\mu\nu\alpha\beta}F^{\mu\nu}F^{\alpha\beta},\,R_{\mu\nu}F^{\mu\alpha}F^\nu_{\,\,\alpha}$ higher dimension operators of $\sqrt{\cor+\cor'}\leq 2,800$\,km.  This is a macroscopic scale of some interest in solar system tests of general relativity. 
This is perhaps surprising (and coincidental) that from Eq.~(\ref{WGCbound}), where we insert the electron mass, the Planck mass, and the fine structure constant, we arrive at such an amusing macroscopic scale. 

In the approximate vacuum of the solar system away from the sun, we have $R_{\mu\nu}\approx0$; so the $\cor'$ term will not play a significant role. While we still have a non-zero $R_{\mu\nu\alpha\beta}$ from the sun; so the $\cor$ term will be essential. The above WGC inequality only refers to the sum, but barring some highly unexpected cancellation\footnote{For example, integrating out the electron at 1-loop leads to  $\Delta\cor'=-13\Delta\cor$ \cite{Drummond:1979pp}. Also, the coefficients should obey $\cor'=-4\cor$ to avoid higher derivative equations of motion. Or we should have $a'=-2a$ to form the Weyl tensor. So in all known examples to us, there is no cancellation of the form $\cor=-\cor'$.}, the WGC bound can be interpreted roughly as $a\lesssim 7.8(1000\,\mbox{km})^2$. So we will be effectively referring to just $\cor$ from now on.
In fact one can go further (see Refs.~\cite{Hamada:2018dde,Alberte:2020bdz}): by performing some field redefinitions, one can take all the $a,\,a',\,a''$ terms to be combined into the form $-a\,W_{\mu\nu\alpha\beta}F^{\mu\nu}F^{\alpha\beta}$, where $W_{\mu\nu\alpha\beta}$ is the Weyl tensor (giving $a'=-2a$ and $a''=a/3$); this makes this a more direct photon-graviton coupling. (Terms of the form $R_{\mu\nu}F^{\mu\alpha}F^\nu_{\,\,\alpha}$ and $RF^2$ are not true graviton-photon couplings in a sense: This is because the Einstein equations tell us to leading order $R_{\mu\nu}$ is determined by $T_{\mu\nu}$, which in turn is determined by a bilinear in $F_{\mu\nu}$ the electromagnetic field strength in vacuum. Upon substitution this means the $a',\,a''$ terms, to leading order, play a redundant role to the terms quartic in $F_{\mu\nu}$, which we are already including.)
Alternatively, one can set $a'=-4a$ and $a''=a$ to achieve two derivative equations of motion \cite{Horndeski:1976gi,Barrow:2012ay,BeltranJimenez:2013btb}.
In either way, the values of $a',\,a''$ terms would be uniquely determined by $a$.

In this work, we are generically interested in placing observational bounds on these corrections. And, by way of motivation, we will wonder what happens if nature saturates or is close to this theoretical bound, i.e.,
$\cor\sim 7.8(1000\,\mbox{km})^2$. While we do not have a good a priori reason to expect nature to be near this theoretical bound, it does provide a test of the WGC. We shall check with existing data whether this is compatible with current solar system tests of gravity.

\section{Altered Light Travel Time}\label{Altered}

One of, if not the, most precise tests of general relativity is the prediction of an increase in the round trip time of light due to a massive object, such as the sun. This is the famous Shapiro time delay \cite{Shapiro:1964uw}.  In the presence of the higher derivative correction $\cor\,R_{\mu\nu\alpha\beta} F^{\mu\nu}F^{\alpha\beta}$ this prediction will be altered. Interesting other works exploring the observational consequences of this operator includes Refs.~\cite{Chen:2016hil,Li:2017zsb,Li:2018jjk,Abbas:2021whh,Zhang:2021hit,Jana:2021lqe,Bamba:2008ja}.

\begin{figure}[t!]
\centering
\includegraphics[width=\columnwidth,height=3.4cm]{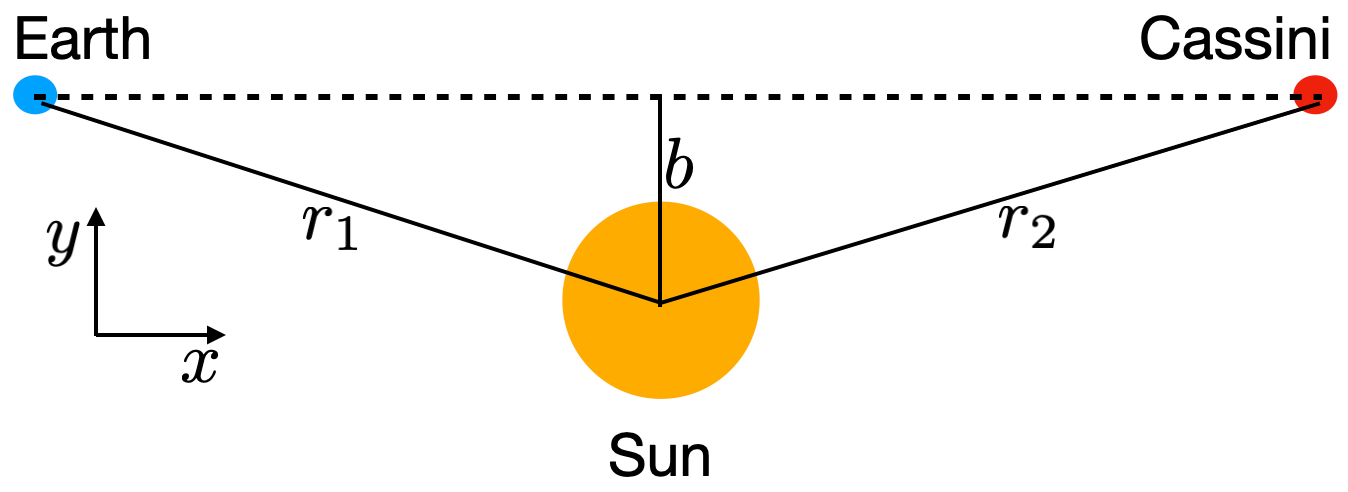}
\caption{Arrangement in solar system: Dashed line is light path of radio waves between earth and Cassini, $b$ is impact parameter, $r_1$ and $r_2$ are the distances from sun to earth and sun to Cassini probe, respectively.}
\label{FigCassiniSunEarth}
\end{figure}

Consider the arrangement in Fig.~\ref{FigCassiniSunEarth}. Here we will be interested in the Cassini probe, which in the year 2002 was near the planet Saturn at just the right time that the light path from Cassini to earth went near the sun; i.e., the light had a small impact parameter $b$ with $b\ll r_1,\,r_2$ (the figure is not to scale).  By sending radio signals from earth to Cassini and back one can test the Shapiro time delay prediction and put constraints on corrections.

In the regime of interest $b\ll r_1,\,r_2$, it is known that the Shapiro time delay for the round trip in general relativity is
\beq
\Delta T_{\mbox{\tiny{GR}}} = {4 G M_{\odot}\over c^3}\,\ln\!\left[4 r_1 r_2\over b^2\right],
\label{Shapiro}\eeq
(where we have reinstated factors of $c$.)
This has been very well measured by the Cassini probe experiment; we shall come to the data in the next section.

Now we wish to compute the correction provided by the higher derivative term $\cor R_{\mu\nu\alpha\beta} F^{\mu\nu}F^{\alpha\beta}$. 
If the light travels along the $x$-axis (dashed line in Fig.~\ref{FigCassiniSunEarth}) from $x_1$ (earth) to $x_2$ (Cassini) and back, the total change in light travel time is
\beq
\Delta T = \Delta T_{\mbox{\tiny{GR}}}+\Delta T_\cor,
\eeq
where the additional term is
\beq
\Delta T_\cor = 
-2\int_{x_1}^{x_2}dx\,\Delta v_\cor/c^2.
\label{Tshift}\eeq
Here $\Delta v_\cor$ represents a (small) correction to the light speed provided by this higher dimension operator in the presence of the sun.

The equation of motion for a photon with wavevector $k_0^\mu$ and polarization vector $\epsilon^{\nu}$ in the geometric optics limit is \cite{Shore:2003zc}
\beq
g^{\mu\nu}k_{\mu}k_{\nu}=8\,\cor\, R_{\mu\nu\alpha\beta}\, k^{\mu}k^{\alpha}\epsilon^\nu\epsilon^\beta.
\eeq
In the solar system, we can use the weak field metric
\beq
ds^2=c^2(1+2\phi_N/c^2)dt^2-(1-2\phi_N/c^2)d{\bf x}^2.
\eeq
Then taking $\phi_N=-G M_{\odot}/r$ from the sun and considering motion of light along the $x$-axis with position vector, (unperturbed) wavevector, and polarization vector
\beq
{\bf x}=(x,b,0),\,\,\,\,\,\,
k_0^\mu=(|k_x|/c,k_x,0,0),\,\,\,\,\,\,\epsilon^\nu=(0,0,\epsilon_y,\epsilon_z)
\eeq
the equation of motion for light becomes 
\beq
g^{\mu\nu} k_\mu k_\nu =-{24\,\pol\,\cor \,G \,M_{\odot} b^2\over (x^2+b^2)^{5/2}c^2}k_x^2,
\eeq
where the polarization structure is captured by the coefficient
\beq
\pol\equiv \epsilon_z^2-\epsilon_y^2.
\eeq
This represents a shift in the speed of light propagation of
\beq
\Delta v_\cor = -{12\,\pol\,\cor \,G \,M_{\odot} b^2\over (x^2+b^2)^{5/2}c}.
\label{vshift}\eeq
We see that this can be positive or negative, depending on the sign of $\cor$ and on the polarization of light:
\bea
&&\pol=\epsilon_z^2-\epsilon_y^2=+1,\,\,\,\mbox{Polarized $\perp$ to solar system},\,\,\,\,\,\,\,\,\,\,\\
&&\pol=\epsilon_z^2-\epsilon_y^2=-1,\,\,\,\mbox{Polarized $\parallel$ to solar system}.
\eea
This means different polarizations of light travel at different speeds; a violation of the equivalence principle. It is known that essentially all higher derivative corrections to general relativity violate the equivalence principle in one form of another, so this is not surprising.

For the case in which $\Delta v_\cor$ is positive, this speed {\em up} should be compared to the speed {\em down} from the usual effect in general relativity (related to the Shapiro time delay).
The maximal value of this speed up is $\Delta v_\cor=12|\cor| G M_{\odot}/(b^3\,c)$, while the maximal value of the speed down within general relativity is $\Delta v_{GR}=-2GM_{\odot}/(bc)$. At this maximum, the effective speed of light is
\beq
v=c\left(1-{2GM_\odot\over bc^2}\left(1-{6|\cor|\over b^2}\right)\right).
\eeq
So in order to avoid superluminality, we need $6|\cor|/b^2<1$. As described earlier, we will be interested in values of $\cor$ in the vicinity of $\cor\sim 10\,(1000\,\mbox{km})^2$, while the smallest impact parameter we will study is $b_{\mbox{\tiny{min}}}=1.6\,R_\odot\approx 1.1\times 10^6\,$km. Hence we have $6|a|/b_{\mbox{\tiny{min}}}^2\sim 5\times 10^{-5}$, ensuring that the behavior is always subluminal (away from the maximum speed correction, the corrections from this new operator are even smaller). We note that other discussions of superluminality in related contexts, includes Refs.~\cite{Daniels:1993yi,Dittrich:1998fy,Novello:1999pg,Shore:2007um,Hollowood:2008kq,Hollowood:2015elj,Goon:2016une,Hertzberg:2017abn,deRham:2020zyh}.

Let us now compute the corresponding time delay. We insert Eq.~(\ref{vshift}) into Eq.~(\ref{Tshift}). Since the integral falls off rapidly at large distances, we can send the endpoints of integration to $x_2\to\infty$ and $x_1\to-\infty$ and we obtain
\beq
\Delta T_\cor={32\,\pol\,\cor\,G\,M_\odot\over b^2 c^3}.
\eeq
This correction should be compared to the standard Shapiro time delay formula in Eq.~(\ref{Shapiro}).

\section{Cassini Probe Data}\label{Data}

Precise measurements can be obtained by comparing the transmitted frequency of light from earth $\nu_{\mbox{\tiny{T}}}$ to the received frequency $\nu_{\mbox{\tiny{R}}}$ after the light reflects off the distant Cassini probe.  The fractional frequency change is defined as
\beq
y={\nu_{\mbox{\tiny{R}}}-\nu_{\mbox{\tiny{T}}}\over \nu_{\mbox{\tiny{T}}}} = -{\Delta T\over dt},
\eeq
where in the second step we noted that frequency is inversely related to period and hence related to the (negative of) the time derivative of light travel time. 

When viewed from the earth, the impact parameter $b$ changes in time as the Cassini probe moved in the solar system relative to the sun. 
Hence we need to consider $b=b(t)$ and take time derivatives accordingly. The frequency shift is then
\beq
y_\cor=-{4GM_\odot\over c^3}{d\over dt}\!\left(\ln\!\left[{4r_1r_2\over b^2}\right]+{8\,\pol\,\cor\over b^2}\right),
\eeq
where we have put a $\cor$ subscript on $y$ to indicate that this is the theoretical prediction that depends on the parameter $a$.
By noting that the much larger scales $r_1,\,r_2$ change much more slowly than $b$, we can ignore differentiating them, giving
\beq
y_\cor=y_{\mbox{\tiny{GR}}}\left(1+{8\,\pol\,\cor\over b^2}\right),
\eeq
where we have factorized for the result in general relativity
\beq
y_{\mbox{\tiny{GR}}}={8GM_\odot\over c^3}{1\over b}{db\over dt}.
\eeq
So we see that the higher derivative correction is a fractional correction of $8\,\pol\,\cor/b^2$.

In July 2002, the Cassini probe was near Saturn, but appeared in the sky from earth with a relatively small impact parameter from the sun with time dependence of
\beq
b(t)=\sqrt{b_{\mbox{\tiny{min}}}^2+v_{\mbox{\tiny{C}}}^2\,t^2},
\eeq
where the minimum impact parameter $b_{\mbox{\tiny{min}}}$ (as mentioned above) and Cassini probe speed $v_{\mbox{\tiny{C}}}$ of
\beq
b_{\mbox{\tiny{min}}}=1.6\,R_\odot,\,\,\,\,\,\,v_{\mbox{\tiny{C}}}\approx 3\,R_\odot/\mbox{day}.
\eeq
The explicit time dependence for the GR result is then
\beq
y_{\mbox{\tiny{GR}}}={8 G M_\odot\over c^3}{v_{\mbox{\tiny{C}}}^2\,t\over b_{\mbox{\tiny{min}}}^2+v_{\mbox{\tiny{C}}}^2\,t^2}.
\eeq

\begin{figure}[t!]
\centering
\includegraphics[width=\columnwidth,height=6.5cm]{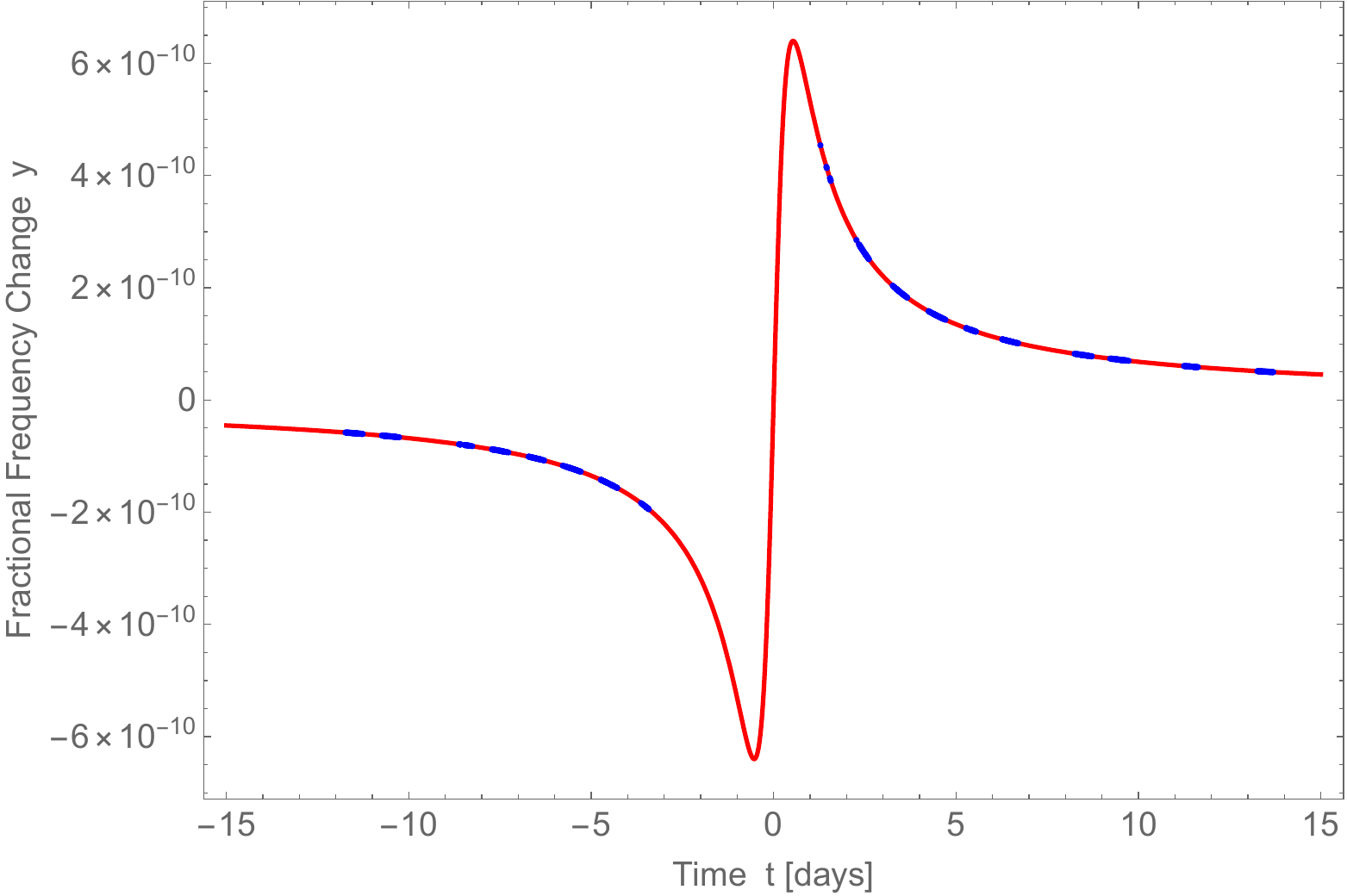}\\
\vspace{0.4cm}
\includegraphics[width=\columnwidth,height=6.5cm]{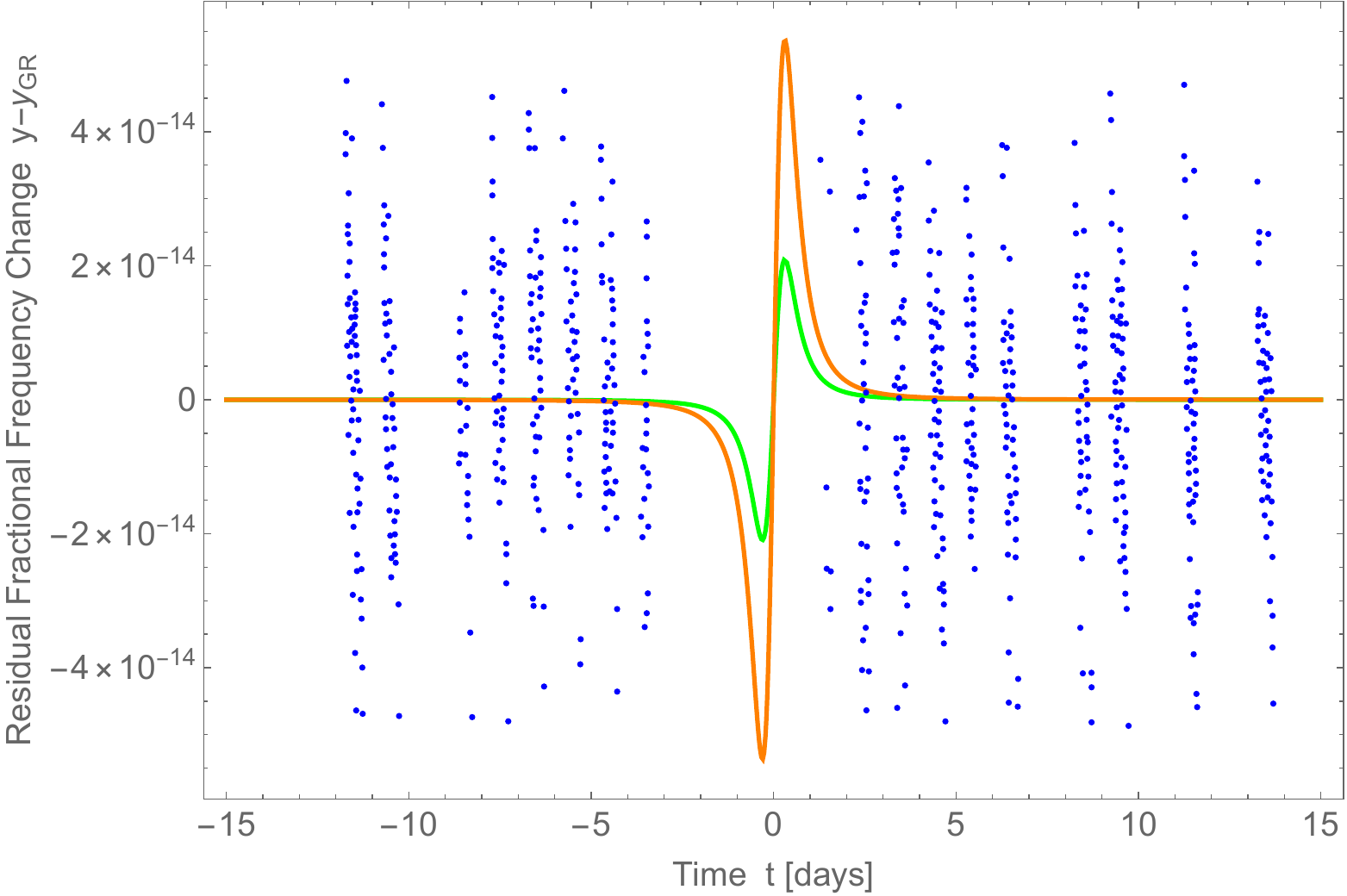}
\caption{Top panel: Fractional frequency change $y$ versus time as measured by an analysis of the Cassini probe data \cite{Bertotti:2003rm}. The red curve is the prediction of general relativity; related to the Shapiro time delay. The blue dots are the data. 
Bottom panel: Fractional frequency change $y$ after subtracting the general relativity prediction to  give the residual $y_{\mbox{\tiny{res}}}=y-y_{\mbox{\tiny{GR}}}$. The green curve is the prediction of the theory with coefficient $\bar\cor=7.8\,(1000\,\mbox{km})^2$,and the orange curve is with coefficient $\bar\cor=20\,(1000\,\mbox{km})^2$.}
\label{FigCassiniData} 
\end{figure}

The data from the Cassini probe was analyzed and processed in Ref.~\cite{Bertotti:2003rm} in order to perform a high precision test of general relativity. The data is given in Fig.~\ref{FigCassiniData}.  The upper plot gives the frequency shift versus time. The solid red curve is the standard prediction from general relativity $y_{\mbox{\tiny{GR}}}$. The blue dots are the data. The lower plot gives the residuals after subtracting out the general relativity value 
\beq
y_{\mbox{\tiny{res,data}}}=y_{\mbox{\tiny{data}}}-y_{\mbox{\tiny{GR}}}.
\eeq
We have also shown on the lower plot the predictions from some representative values of the higher derivative coeffcient $\cor$. Since there is polarization dependence, we have defined $\bar{a}=\pol\,\cor$. The green curve is for $\bar{a}=7.8\,(1000\,\mbox{km})^2$ and the orange curve is for $\bar{a}=20\,(1000\,\mbox{km})^2$. Although it is hard to notice a big disagreement compared to the data in the central region of interest, since there are so many data points, we can still significantly constrain $\cor$ from this.

\section{Likelihood Results}\label{Likelihood}

We wish to use the data to put constraints on the coefficient $\cor$. We test values of $\cor$ against the null hypothesis $\cor=0$ (which is general relativity).

We run a likelihood analysis by noting that each data point is Gaussian distributed, with a central value shown in Fig.~\ref{FigCassiniData}, and a standard deviation of
\beq
\sigma_y=1.4\times 10^{-14}.
\eeq
Treating each data point as independent, the relative likelihood is
\beq
L=\prod_j^N\,{\exp\!\left[-(y_{\mbox{\tiny{res,data}}}-y_{\mbox{\tiny{res}},\cor})^2/(2\sigma_y^2)\right]\over \exp\!\left[-(y_{\mbox{\tiny{res,data}}})^2/(2\sigma_y^2)\right]},
\eeq
where $y_{\mbox{\tiny{res,data}}}=y_{\mbox{\tiny{data}}}-y_{\mbox{\tiny{GR}}}$ is the central value of the data and $y_{\mbox{\tiny{res}},\cor}=y_\cor-y_{\mbox{\tiny{GR}}}$ is the value of the theory with parameter $\cor$. The denominator normalizes this to the null hypothesis of $\cor=0$. The product is over each data point labelled $j=1,\ldots,N$.

\begin{figure}[t!]
\centering
\includegraphics[width=\columnwidth,height=6.5cm]{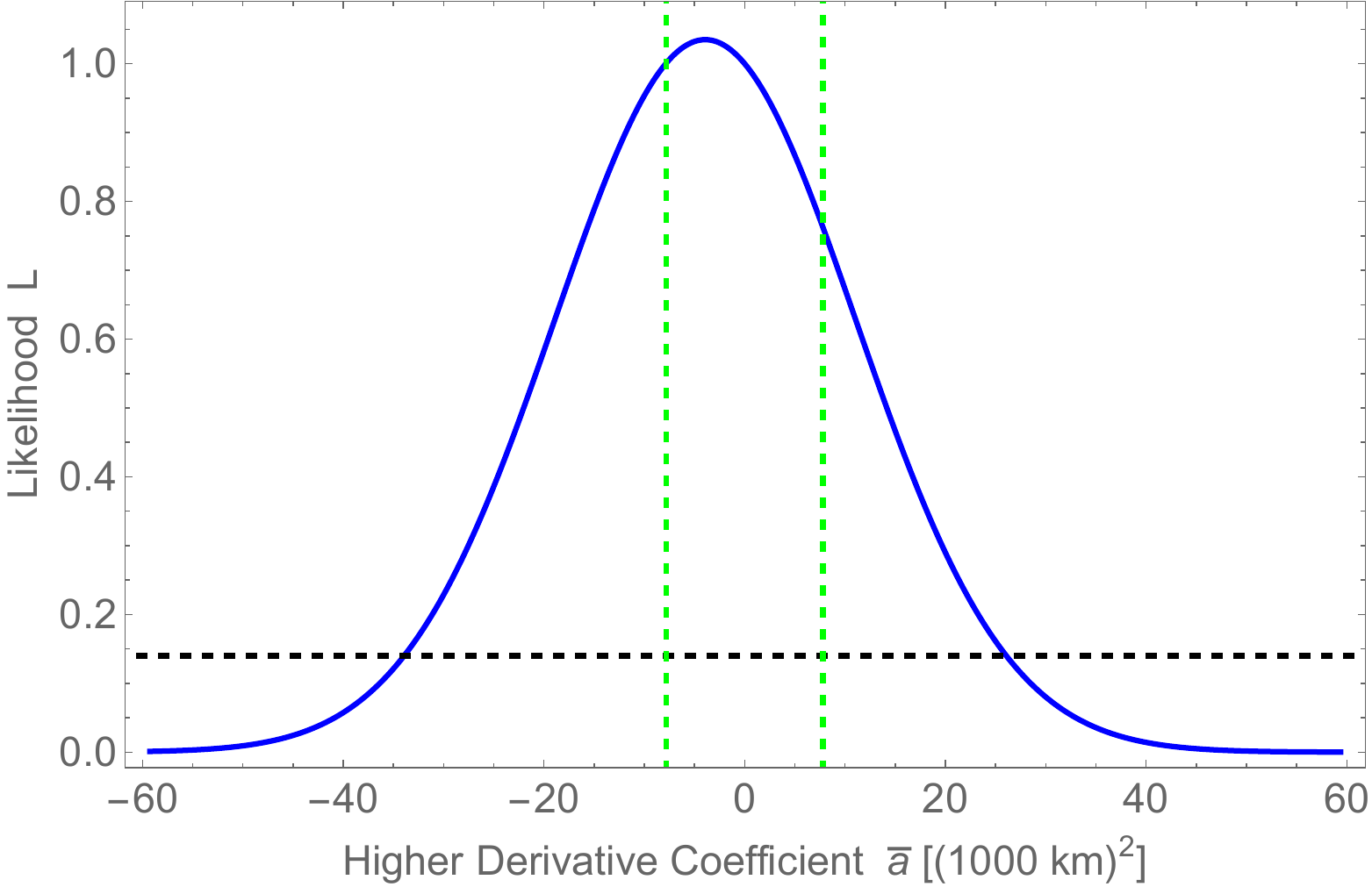}
\caption{Likelihood $L$ versus parameter $\bar{a}=\pol\,\cor$. Dashed line intersects the 2-sigma interval. The vertical dashed green lines are the bounds indicated by the WGC.}
\label{FigLikelihood} 
\end{figure}

We have run this analysis by multiplying over the set $j$ with $N=724$ the number of available data points that we had access to. We have examined a range of values of the parameter $\cor$. Our results for the likelihood $L$ are given in Fig.~(\ref{FigLikelihood}). 

Recall that the relation between $a$ and $\bar{a}$ is that $\bar{a}=\varepsilon a$. And the value of $\varepsilon$ is $\varepsilon=+1$ for polarization perpendicular to the plane of Cassini-earth sun, and $\varepsilon=-1$ for polarization parallel parallel to the plane of Cassini-earth-sun. These are orthogonal linear polarizations which are normal modes of this system. One can indeed have $\varepsilon$ between -1 and 1 too, but not for the normal modes. So such other modes are superpositions of the normal modes, and the other values will have non-trivial time dependence. 
In Fig.~(\ref{FigLikelihood}), the constraints (blue curve) pertain to the variable $\bar{a}$ as the correction to the Shapiro delay is proportional to this. The vertical green lines indicate the WGC bounds, which are proportional to $a$. But since $|\varepsilon|\leq 1$, we can say that the WGC bound can live {\em inside} this region, so these vertical lines are  useful reference values.

Since the theoretical prediction $y_{\mbox{\tiny{res,\cor}}}$ is proportional to $\cor$, the likelihood $L$ is itself a Gaussian in $\cor$. We can write this as
\beq
L\propto\exp\left[-(\bar{\cor}-\bar{\cor}_p)^2/(2\sigma_a^2)\right]
\eeq
and we have found the mean and standard deviation to be
\bea
&&\bar{\cor}_p\approx-4\,(1000\,\mbox{km})^2\\
&&\sigma_a\approx 15\,(1000\,\mbox{km})^2
\eea
We see that the data has a very mild preference for negative values of $\bar{\cor}$, with peak value of 
$\bar{\cor}_p=-4\,(1000\,\mbox{km})^2$. 
But this is not statistically significant as the null hypothesis of $\cor=0$ is within the 1 sigma region. 

We report a constraint on the parameter $\cor$ at the 2-sigma ($\approx 95\%$) confidence level. Again absorbing the polarization dependence into $\bar{\cor}=\pol\,\cor$, we have
\beq
-34\,(1000\,\mbox{km})^2<\bar{\cor}<26\,(1000\,\mbox{km})^2. 
\eeq
This is indicated by where the blue likelihood curve $L$ intersects the horizontal dashed line in the figure. Since both polarizations should be present in the radio waves sent back and forth from earth to the Cassini probe, then the smallest side of this range gives the corresponding bound on $\cor$
\beq
|\cor|<26\,(1000\,\mbox{km})^2.
\eeq

In the figure, we have also indicated the values $\bar{a}= \pm 7.8(1000\,\mbox{km})^2$ as the vertical dashed green lines. Assuming $\cor+\cor'$ is positive, then we should remain within the range by the WGC $|\cor|\lesssim7.8(1000\,\mbox{km})^2$ or one of the polarizations would be inconsistent. (If $\cor+\cor'$ is negative, we have no direct bound from WGC).

Overall the prediction of the WGC is compatible with the existing data. Interestingly, the WGC bound and the observational bound are close with this existing data set.

\section{Discussion}\label{Discussion}

We have found an observational bound on the $R_{\mu\nu\alpha\beta}F^{\mu\nu}F^{\alpha\beta}$ coefficient of $|\cor|<26\,(1000\,\mbox{km})^2$. An important point is that the light propagation is polarization dependent. Ideally one would separate out the two polarizations and analyze each individually. However the data is not currently presented in this form. Instead we have exploited the fact that each individual data point has not shown a {\em splitting} into pairs (one raised and one lowered), or an enhanced variance, as would be expected if the higher dimension operator is altering the propagation, and hence we can still infer a bound on $\cor$. 

Moreover, consider again the 2 eigenstates for this system: $\varepsilon=+1$ for polarization perpendicular to the plane of Cassini-earth sun and $\varepsilon=-1$ for polarization parallel to the plane of Cassini-earth-sun. Suppose we have one of these, say $\varepsilon=-1$ for polarization parallel to the plane to Cassini-earth sun, we anticipate that after reflecting off a mirror it will remain parallel. i.e., if the electric field is oscillating in such a plane, we see no reason it should rotate after hitting a mirror and returning. 
It can pick up a phase shift, but that does not alter the corrected time delay we are computing here. It will be reinforced in the round trip for a given polarization. However, if there is a mixture of polarizations, it is more subtle and possible cancellations can occur.

Our statistical analysis is overall a first pass, given that we do not have explicit polarization data.
It would be useful in future work to search explicitly for a possible deviation in the behavior of each polarization. 
This may lead to a reduction in precision, or alternatively,  a significantly improved bound (or a possible detection).

In this work, since we have been considering large values of $\cor$, in the vicinity of $\cor\sim10\,(1000\,\mbox{km})^2$. 
Such a large coupling between the photon and the graviton, would lead to a bad breakdown of perturbativity if taken seriously to high energies. 
For example, consider a pair of photons scattering via graviton exchange. Using this new operator, the scattering amplitude at energy $E$ would be $\mathcal{M}\sim \cor^2E^6/\mpl^2$. This violates the unitarity bound $\mathcal{M}\lesssim 1$ at the scale $E_{\mbox{\tiny{UV}}}\sim(\mpl/\cor)^{1/3}$. 
Using $\cor\sim 10\,(1000\mbox{km})^2$ this would be $E_{\mbox{\tiny{UV}}}\sim 2\,$eV. Furthermore, one might anticipate new physics already entering at the extremely low energy scale $1/\sqrt{|\cor|}$; although the need for new physics at this scale is less clear. One can potentially remove this scale from playing the role of a cut off by imposing $\cor'=-4\cor=-4\cor''$, which prevents the equations of motion from being higher order in derivatives, avoiding ghost modes.
In any case, even the above unitarity cut off at $E_{\mbox{\tiny{UV}}}\sim 2$\,eV is a rather low cut off for a useful effective field theory. Since we have yet to see new physics in the electromagnetic-gravitational sector, it hints that this construction is implausible. However, since it involves gravity, one can have an open mind to the possibility that some non-trivial physics in the gravitational sector provides the UV completion. Though we do not have a concrete example of this. 

Further work can include considering other observations to constrain the term $\cor R_{\mu\nu\alpha\beta}F^{\mu\nu}F^{\alpha\beta}$. But also to consider observations to constrain $a'R_{\mu\nu}F^{\mu\alpha}F^\nu_{\,\,\alpha}$. The latter requires considering regions in which there is matter, so that the Ricci tensor is non-zero. This is important since the WGC is actually sensitive to the sum $\cor+\cor'$. Alternatively, one could try to sharpen the argument that $\cor'=-4\cor$ to avoid ghosts in the UV, which would link the coefficients. Or use the Weyl tensor coupling, which links the coefficients as $a'=-2a$, and provides a more direct probe of graviton-photon coupling.


\section*{Acknowledgments}
M.~P.~H.~ is supported in part by National Science Foundation grant PHY-2310572. 
We thank the VERSE program at Tufts University for support.
We thank John Donoghue, Ben Heidenreich, Luciano Iess, Paolo Tortora, and especially Sera Cremonini for helpful discussion.


\end{document}